\documentclass[reprint,
superscriptaddress,
preprintnumbers,
 amsmath,amssymb,
 aps,
prl,longbibliography,
]{revtex4-2}
\usepackage{CJKutf8}
\usepackage{multirow}
\usepackage{natbib}
\usepackage{mathtools}
\usepackage{xcolor}
\usepackage{graphicx}
\usepackage{dcolumn}
\usepackage{bbm}
\usepackage{hyperref}

\DeclareMathAlphabet\mathbfcal{OMS}{cmsy}{b}{n}

\newcommand{\ra}{\rangle}

\newcommand{\p}{\partial}
\newcommand{\ve}{\varepsilon}
\newcommand{\la}{\langle}

\newcommand{\eq}{\begin{equation}}
\newcommand{\eqe}{\end{equation}}
\newcommand{\eqa}{\begin{eqnarray}}
\newcommand{\eqae}{\end{eqnarray}}
\newcommand{\nn}{\nonumber}
\newcommand{\bn}{\begin{enumerate}}
\newcommand{\en}{\end{enumerate}}
\def\beq#1\eeq{\begin{align}#1\end{align}}

\newcommand{\eqc}[1]{(\ref{#1})}

\def\CA{{\mathcal A}}

\def\CC{{\mathcal C}}

\def\CE{{\mathcal E}}

\def\CH{{\mathcal H}}

\def\ve{\varepsilon}

\def\half{\frac{1}{2}}

\newcommand{\bfig}{\begin{figure}}
\newcommand{\efig}{\end{figure}}

\def\abs#1{{\left| #1 \right|}}

\def\bl#1\el{\begin{align} #1 \end{align}}
\def\bg#1\eg{\begin{gather} #1 \end{gather}}

\def\bld#1\eld{\begin{aligned} #1 \end{aligned}}
\def\bgd#1\egd{\begin{gathered} #1 \end{gathered}}

\newcommand{\bra}[1]{\langle{#1}|}
\newcommand{\ket}[1]{|{#1}\rangle}

\newcommand{\sbra}[1]{ [{#1} |}

\renewcommand{\tt}{\texttt}

\newcommand{\RN}[1]{%
  \textup{\uppercase\expandafter{\romannumeral#1}}%
}

\newcommand{\be}{\begin{equation}}
\newcommand{\ee}{\end{equation}}
\newcommand{\ba}{\begin{align}}
\newcommand{\ea}{\end{align}}
\newcommand{\bi}{\begin{itemize}}
\newcommand{\ei}{\end{itemize}}

\let\a=\alpha \let\b=\beta \let\g=\gamma  \let\e=\epsilon
    \let\k=\kappa
\let\l=\lambda \let\m=\mu \let\n=\nu \let\x=\xi \let\p=\pi \let\r=\rho 
\let\s=\sigma \let\t=\tau  \let\f=\phi  
\let\w=\omega   \let\G=\Gamma \let\D=\Delta  \let\L=\Lambda

\makeatletter
\newcommand*{\Rom}[1]{\expandafter\@slowromancap\romannumeral #1@}
\makeatother

\makeatletter
\newcommand*{\rom}[1]{\expandafter\romannumeral #1}
\makeatother

\let\nn=\nonumber

\def\beq#1\eeq{\begin{align}#1\end{align}}

\begin{document}
\begin{CJK*}{UTF8}{mj}
\preprint{QMUL-PH-20-31}
\title{Quantum corrections to tidal Love number for Schwarzschild black holes
}
\author{Jung-Wook Kim (김정욱)}
\email{jung-wook.kim@qmul.ac.uk}
\affiliation{Centre for Research in String Theory, School of Physics and Astronomy,\\Queen Mary University of London, Mile End Road, London E1 4NS, United Kingdom}
\author{Myungbo Shim (심명보)}
\email{mbshim1213@khu.ac.kr}
\affiliation{Department of Physics and Research Institute of Basic Science,\\Kyung Hee University, Seoul 02447, Korea}

\begin{abstract}
A sum rule for tidal Love number is derived from quantum field theory computations, which relates tidal susceptibility of a spinless body to transition rates of its single graviton emission processes. An analogous sum rule for electromagnetism is given as an example, which is substantiated by comparing to the solved problem of the hydrogen atom. Based on the semiclassical Hawking radiation spectrum, a finite nonvanishing value for quantum corrections to the Love number of Schwarzschild black holes in general relativity is computed using the sum rule, which is known to classically vanish.
\end{abstract}

\maketitle
\end{CJK*}

\section{Introduction}
At sufficiently large scales, any finite-sized system can be considered as a point particle. The point particle approximating a macroscopic system does not need to be classical; some of the recent progress in the gravitational two-body problem~\cite{Bern:2019nnu,Bern:2019crd,Guevara:2019fsj,Chung:2020rrz,Cheung:2020sdj} relied on treating the gravitating bodies as point particles of quantum field theory~(QFT), where advances in tools for computing QFT amplitudes~\cite{Bern:1994cg,Bern:1994zx,Forde:2007mi,Kawai:1985xq,Britto:2004apm, Britto:2005fq,Bern:2008qj,Bern:2010ue,Bern:2010yg,Arkani-Hamed:2017jhn} played a crucial role. This paper aims to show that treating a macroscopic spinless body as a quantum point particle gives an interesting insight into the relation between a body's tidal susceptibility and its quantum spectrum,
and this relation can be used to infer quantum corrections to the tidal susceptibility of Schwarzschild black holes (SBHs) of asymptotically flat spacetime in general relativity.

Tidal Love numbers parametrize susceptibilities of the body under tidal influences~\cite{doi:10.1098/rspa.1909.0008,poisson_will_2014}, which are analogs of electric and magnetic susceptibilities in electromagnetism. They have gained interest as tidal interactions and are expected to leave recognizable imprints on the waveforms of gravitational waves, which can be detected by ground-based interferometers such as LIGO~\cite{Bildsten:1992my,Flanagan:2007ix,Hinderer:2009ca,Damour:2012yf,Abbott:2018wiz,Abbott:2018exr}. Their measurement is important for studying neutron stars, as they provide model-independent constraints on their physics~\cite{Yagi:2013bca,Yagi:2013awa}. 

In this paper, a matching computation between amplitudes constructed from point particle effective action and amplitudes constructed from modern QFT techniques will be used to derive constraints on susceptibilities of a body. The worldline action for an effective point particle with worldline operators parametrizing the tidal susceptibilities can be written as~\cite{Goldberger:2007hy,Damour:2009vw,Damour:2009wj}
\bl
\bld
S &= \int (- m \sqrt{u^2} + \frac{C_E}{\sqrt{u^2}^3} E_{\m\n} E^{\m\n} + \frac{C_{B}}{\sqrt{u^2}^3} B_{\m\n} B^{\m\n} ) d\s \,,
\eld \label{eq:EffActGrav}
\el
where $u^2 = g_{\m\n} u^\m u^\n$ and 
\beq
u^\m = \frac{d x^\m(\s)}{d \s}\,,
\eeq
where $E_{\m\n}$($B_{\m\n}$) are electric(magnetic) components of the Weyl tensor, defined as~\footnote{The convention is $R^{\m}_{~\n\l\s} = \partial_\l \G^{\m}_{\s\n} + \cdots$ and mostly negative signature.}
\beq
E_{\m\n} = W_{\m\a\n\b} u^\a u^\b \,,
&& B_{\m\n} = \frac{1}{2} \e_{\a\b\g\m} W^{\a\b}_{\phantom{\a\b}\delta\n} u^\g u^\delta \,.
\eeq
In the weak field limit, $E_{ij} = - \partial_i \partial_j \Phi$ where $\Phi$ is the Newtonian potential. These worldline operators will be converted to the amplitude contributions \eqc{eq:GravComptonEff}.

As noted in Ref.\cite{Kim:2020cvf}, the measurement of tidal susceptibilities can be modeled as a process of absorbing a low-energy graviton and transitioning to an intermediate state of differing rest mass, before transitioning back to the original state by reemitting another graviton with the same energy. This process is described as the amplitude \eqc{eq:ComptonGrav}, which was constructed using modern QFT techniques. The matching computation between \eqc{eq:ComptonGrav} and \eqc{eq:GravComptonEff} results in the following relation between electric Love number $C_E$ and single graviton emission rates
\bl
C_E &=  \frac{1}{8G} \left[ \sum_{m'>m} \frac{\G_{m' \to mh}}{(\D m)^6} - \sum_{m'<m} \frac{\G_{m \to m'h}}{(\D m)^6} \right] \,, \label{eq:gravdecay2pol3} 
\el
where $m$ is a scalar particle, $m'$ is a spin-2 particle of mass $m'=m+\D m$, and $\G_{m' \to mh}$($\G_{m \to m'h}$) is the total transition rate of particle $m'$($m$) decaying into particle $m$($m'$) by emitting a graviton. Related attempts for writing a sum rule can be found in refs.\cite{Rothstein:2014sra,Brustein:2020tpg,Brustein:2021bnw} where only the first term on the right of \eqc{eq:gravdecay2pol3} was considered; these approaches treated as if the system is in its ground state, so the particle cannot transition to a state with lower energy. A similar relation for the first term in a classical context has been given in Ref.\cite{Chakrabarti:2013xza} as well.

This sum rule can be used to compute quantum corrections to the tidal Love number of SBHs by treating them as scalar particles~\footnote{The timescale under consideration is extremely short compared to the typical evaporation time of black holes, so black holes can be practically considered as stable particles.}. Assuming the semiclassical Hawking radiation (HR) spectrum fully captures the numerators of \eqc{eq:gravdecay2pol3}, the computed finite correction with $G$, $c$, and $\hbar$ restored is (see \eqc{eq:HRlovePlanck} for dimensionless Love number $k_2$)
\bl
C_{E}^{\text{HR}} &=  \frac{G^3 M^3 \hbar}{135 \pi c^9} \,. \label{eq:HRlove} 
\el
To the best of authors' knowledge, this is the first computation on quantum corrections to tidal Love numbers of black holes within general relativity without any undetermined parameters. 
From the perspective that vanishing of the Love number is related to the no-hair property of black holes~\cite{Gurlebeck:2015xpa}, the result may be interpreted as an indication that SBHs of general relativity carry quantum hair.

To understand the significance of the result \eqc{eq:gravdecay2pol3}, it is helpful to understand the derivation of its electromagnetic cousin and how it relates to an analogous computation in quantum mechanics, using s-orbitals of the hydrogen atom as explicit examples. Natural units $\hbar=c=1$ will be used throughout the paper unless explicitly stated.


\section{Susceptibility sum rule in QED}
The on-shell three-point amplitude of decay process $m' \to m\g$ is fixed by kinematics~\cite{Arkani-Hamed:2017jhn}.
\bl
\bld
\CA_{m'm\g}^+ = \frac{e \CC}{m} [\mathbf{1}3]^2 \,,
&& \CA_{m'm\g}^- = \frac{e \CC}{m} \la \mathbf{1}3 \ra^2 \,,
\eld \label{eq:QED3ptDecay}
\el
where $\g$ is a photon, $m'$ is a spin-1 massive particle, and $m$ is a massive scalar, with labels indicating their respective masses. This is the amplitude with the lowest power of $\omega$ dependence for the decay rate. The conventions for massive spinor-helicity formalism follows that of Ref.\cite{Chung:2018kqs}. The coefficient $\CC$ is a dimensionless parameter determining the total decay rate $\G_{m' \to m\g}$.
\bl
\bld
\G_{m' \to m\g} &= 4 \pi \times \frac{\omega}{32 \pi^2 m^2} \sum_{m',\g} \abs{\CA_{m'm\g}}^2 
\\ &= \left( \frac{e\CC}{m} \right)^2 \frac{\omega^3}{\pi} \,,
\eld \label{eq:decayrateQFT}
\el
where the sum is over polarizations of $m'$ and $\g$. The nonrelativistic approximation $m' = m+\D m \simeq m$ was used in appropriate places to simplify the expression. Note that the decay rate $\G_{m \to m'\g}$ relevant for $m'<m$ is given by the same equation \eqc{eq:decayrateQFT} with $\omega = \D m$ changed to $\omega = - \D m$. Preparing a pair of these amplitudes and gluing them along the $m'$ leg following the procedure of Ref.\cite{Arkani-Hamed:2017jhn} results in the four-point helicity amplitudes
\bl
\bld
\CA_{m\g \g m}^{++} &= - \left( \frac{e\CC m'}{m} \right)^2 [23]^2 \left[ \frac{1}{s - m'^2} + \frac{1}{u-m'^2} \right] \,,
\\ \CA_{m\g \g m}^{+-} &= - \left( \frac{e\CC}{m} \right)^2 \sbra{2}p_1\ket{3}^2 \left[ \frac{1}{s - m'^2} + \frac{1}{u-m'^2} \right] \,,
\eld \label{eq:Compton}
\el
where the Mandelstam variables are defined as~\footnote{Conforming with the standards of amplitudes literature, all-outgoing convention is used.}
\bl
s = (p_1 + p_2)^2 \,,\, t = (p_1 + p_4)^2 \,,\, u = (p_1 + p_3)^2 \,.
\el
In the nonrelativistic low-energy limit the amplitudes become ``contact" interactions.
\bl
\bld
\CA_{m\g \g m}^{++} 
&=  \left( \frac{e\CC}{m} \right)^2 \left[ \frac{2}{\D m} + \frac{4}{m} \right] \frac{m [23]^2}{2} \,,
\\ \CA_{m\g \g m}^{+-} 
&=  \left( \frac{e\CC}{m} \right)^2 \frac{2}{\D m} \frac{\sbra{2}p_1\ket{3}^2}{2m} \,,
\eld \label{eq:ComptonContact1}
\el
where the nonrelativistic low-energy limit is defined as the limits
\bl
\abs{t} , \abs{s - m^2} \ll \abs{m'^2 - m^2} \,,\, \D m / m \ll 1 \,.
\el

The above ``contact" interactions can be compared with amplitudes arising from the following effective one-particle worldline action.
\bl
\bld
S &= \int (- m - q A_\m u^\m - \frac{\a}{2} E_\m E^\m - \frac{\chi}{2} B_{\m} B^{\m} ) d\t \,,
\eld \label{eq:EffAct}
\el
where
\bg
u^\m = \frac{d x^\m(\t)}{d \t} \,,\, E^\m = F^{\m\n} u_\n \,, B^{\m} = \frac{1}{2} \e^{\m\a\b\g} u_\a F_{\b\g} \,,
\eg
so that in the rest frame of the particle $E^\m = (0, \vec{E})$ is the electric field felt by the particle and $B^{\m} = (0,\vec{B})$ is the magnetic field felt by the particle. From the classical mechanics and electrostatics relations
\bl
S = \int (T- V) dt \,, && V = - \vec{p} \cdot \vec{E} - \vec{\m} \cdot \vec{B} \,,
\el
where $\vec{p}$ is the electric dipole moment and $\vec{\m}$ is the magnetic dipole moment, the expectation value for the dipole moments will be given by the relations
\bl
\bld
\la \vec{p} \ra = \left\langle \frac{\partial S}{\partial \vec{E}} \right\rangle = \a \la \vec{E} \ra \,,
&& \la \vec{\m} \ra = \left\langle \frac{\partial S}{\partial \vec{B}} \right\rangle = \chi \la \vec{B} \ra \,.
\eld \label{eq:dipolesource}
\el
The action can be converted to amplitudes using the substitution rules introduced in Ref.\cite{Chung:2018kqs}
\bl
\bld
\partial_\m A_\n \to - i k_\mu \ve_\n (k) \,,
&& u^\m \to p_1^\m / m \,,
\eld
\el
where $\ve_\m (k)$ is the polarization vector of the photon with momentum $k^\m$. This results in the on-shell helicity amplitudes
\bl
\bld
\CA_{m\g \g m}^{++} &= (\a - \chi) \frac{N [23]^2}{4} \,,
\\\CA_{m\g \g m}^{+-} &= (\a + \chi) \frac{N \sbra{2} p_1 \ket{3}^2}{4m^2} \,,
\eld \label{eq:ComptonEff1}
\el
up to an overall constant $N$. This constant can be fixed by matching to the on-shell three-point amplitude of scalar QED
\bl
\bld
\CA_{mm\g}^{\text{Eff. Act.}} &= - q N \ve_\m (p_1 / m)^\m \,,
\\ \CA_{\bar{\f} {\f} \g}^{\text{Scalar QED}} &=  q \ve_\m (p_2 - p_1)^\m \,.
\eld
\el
Comparing the two fixes the proportionality constant as $N = 2m$, yielding
\bl
\bld
\CA_{m\g \g m}^{++} &=  (\a - \chi) \frac{m [23]^2}{2} \,,
\\ \CA_{m\g \g m}^{+-} &=  (\a + \chi) \frac{\sbra{2} p_1 \ket{3}^2}{2m} \,.
\eld \label{eq:ComptonEff} 
\el
Matching \eqc{eq:ComptonContact1} to \eqc{eq:ComptonEff} determines the susceptibilities as
\bl
\a = \left( \frac{e\CC}{m} \right)^2 \left[ \frac{2}{\D m} + \frac{2}{m} \right] \,,\, \chi = - \left( \frac{e\CC}{m} \right)^2 \frac{2}{m} \,. \label{eq:susceptQFT} 
\el
The above result gives an explanation for the suppression of magnetic susceptibility compared to its electric counterpart; $\chi / \a \simeq \D m / m$. Matching \eqc{eq:decayrateQFT} to \eqc{eq:susceptQFT} in the nonrelativistic limit yields the relation
\bg
\frac{\G_{m' \to m\g}}{\a} =  \frac{\omega^3 \D m}{2\pi} =  \frac{\omega^4}{2\pi} \,. \label{eq:decay2pol} 
\eg
This relation becomes a sum over $m'$ when more than one channel exists.
\bl
\a &=  2\pi \sum_{m'} \frac{\G_{m' \to m\g}}{(\D m)^4} \,. \label{eq:decay2pol2} 
\el
Including the contributions from $m > m'$, i.e., incorporating the decay channels $m \to m' \g$ with $\D m = - \omega$, yields the electromagnetic version of the sum rule \eqc{eq:gravdecay2pol3}.
\bl
\a &=  2\pi \left[ \sum_{m'>m} \frac{\G_{m' \to m\g}}{(\D m)^4} - \sum_{m'<m} \frac{\G_{m \to m' \g}}{(\D m)^4} \right] \,. \label{eq:decay2pol3} 
\el

\section{An example: hydrogen atom}
The hydrogen atom is an exactly solvable model which can be used to confirm the validity of \eqc{eq:decay2pol3}. The quantum mechanical Hamiltonian of the system is
\beq
\CH_{0}=\frac{p^{2}}{2m_{e}}-\frac{e^{2}}{4\p r},&&\CH_{Stark}=eEz,
\eeq
where $\CH_{Stark}$ is treated as a perturbing potential. For a generic s-orbital state of principal quantum number $n$, the induced dipole moment of the hydrogen atom at leading perturbation order \cite{Landau:1991wop} is given by
\beq
\a_{n}=\sum_{n'\neq n}\a_{n}^{(n')},&&\a_{n}^{(n')}\equiv 2e^{2}\frac{\abs{\bra{n',1,0}z\ket{n,0,0}}^{2}}{\CE_{n'}-\CE_{n}},\label{eq:QMpolar} 
\eeq
where $\CE_{n}$ is the energy of orbital states having $n$ as the principal quantum number and $\a_{n}^{(n')}$ is the contribution to the polarizability coming from overlap of $n'$ and $n$ orbitals. On the other hand, the spontaneous emission rates \cite{sakurai1967advanced} involving the same s-orbital state are given by
\beq
\G_{(n',n)}&\equiv\sum_{m\in\{0,\pm1\}}\frac{\w_{(n,n')}^{3} e^{2}}{\p\hbar c^3}\frac{\abs{\bra{n',1,m}\vec{x}\ket{n,0,0}}^{2}}{3}\,,\label{eq:decayrateQM}
\eeq
where $\w_{(n,n')}=|\CE_{n'}-\CE_{n}|/\hbar$. Comparing the matrix elements, (\ref{eq:QMpolar}) and (\ref{eq:decayrateQM}) can be combined into the relation
\bl
\a_{n} &= 2\p c^{3} \left(\sum_{n'>n}\frac{\G_{(n,n')}}{\w^{4}_{(n,n')}}-\sum_{n'<n}\frac{\G_{(n,n')}}{\w^{4}_{(n,n')}}\right) \,, \label{eq:decayQM} 
\el
which is exactly \eqc{eq:decay2pol3} when natural units $c =  1$ are used.

\section{Susceptibility sum rule in gravity}
The analogs of \eqc{eq:QED3ptDecay} for gravity are also fixed by kinematics~\cite{Arkani-Hamed:2017jhn,Kim:2020cvf}
\bl
\bld
\CA_{m'm h}^{+} = \frac{\k \tilde\CC}{m^2} [\mathbf{1}3]^4 \,,
&& \CA_{m'm h}^{-} = \frac{\k \tilde\CC}{m^2} \la \mathbf{1}3 \ra^4 \,,
\eld
\el
where $m'$ is a massive spin-2 particle, $\k = \sqrt{32 \pi G}$ is the gravitational coupling and $\tilde\CC$ is a dimensionless parameter for the total decay rate $\G_{m' \to mh}$.
\bl
\bld
\G_{m' \to mh} &= 4\pi \times \frac{\omega}{32 \pi^2 m^2} \sum_{m',h} \abs{\CA_{m'mh}}^2
\\ &=\left( \frac{\k \tilde\CC}{m} \right)^2 \frac{4 \omega^5}{\pi} \,.
\eld \label{eq:GravDR}
\el
Again, nonrelativistic approximation was used in appropriate places to simplify the expression. The \emph{modifications} of four-point helicity amplitudes~\footnote{While it is possible to have charge neutral particles in electromagnetism, this is impossible in gravity as gravitons couple to all forms of energy. Therefore \eqc{eq:ComptonGrav} must be understood as corrections to the gravitational Compton amplitudes given in refs.\cite{Arkani-Hamed:2017jhn, Caron-Huot:2018ape, Chung:2018kqs, Johansson:2019dnu, Chung:2019duq, Kim:2020cvf}.} are given as
\bl
\bld
\D \CA_{mhhm}^{++} &= - \left( \frac{\k \tilde\CC m'^2}{m^2} \right)^2 [23]^4 \left[ \frac{1}{s - m'^2} + \frac{1}{u-m'^2} \right] \,,
\\ \D \CA_{mhhm}^{+-} &= - \left( \frac{\k \tilde\CC}{m^2} \right)^2 \sbra{2}p_1\ket{3}^4 \left[ \frac{1}{s - m'^2} + \frac{1}{u-m'^2} \right] \,,
\eld \label{eq:ComptonGrav}
\el
where the same definition for the Mandelstam variables were used. The nonrelativistic low-energy expansion of \eqc{eq:ComptonGrav} gives the ``contact" interactions analogous to \eqc{eq:ComptonContact1}
\bl
\bld
\D \CA_{mhhm}^{++} &= \left( \frac{\k \tilde\CC}{m} \right)^2 \left[\frac{1}{\D m} + \frac{4}{m} \right] m [23]^4\,,
\\ \D \CA_{mhhm}^{+-} &= \left( \frac{\k \tilde\CC}{m} \right)^2 \frac{1}{\D m} \frac{\sbra{2}p_1\ket{3}^4}{m^3} \,.
\eld \label{eq:GravComptonContact1}
\el
The analog of \eqc{eq:ComptonEff1} can be obtained from the effective action \eqc{eq:EffActGrav} by expanding the metric around flat space $g_{\m\n} = \eta_{\m\n} + \k h_{\m\n}$. Imposing the normalization condition $\eta_{\m\n} u^\m u^\n = 1$, the relevant leading contributions of the effective point particle action (\ref{eq:EffActGrav}) are
\bl
\bld
S &\simeq \int ( - m - \frac{\k m}{2} h_{\m\n} u^\m u^\n 
\\ &\phantom{asdfasdf}+ C_E E_{\m\n} E^{\m\n} + C_B B_{\m\n} B^{\m\n} ) d\s\,.
\eld
\el
The relation between the induced mass quadrupole moment $Q_{ij} \sim \int d\r (x^i x^j - \frac{x^2 \delta^{ij} }{3})$ and the tidal background $E_{ij} = -\partial_i \partial_j \Phi$ in the Newtonian limit is given as
\bl
V_{int} = - \half Q_{ij} E_{ij} && \Rightarrow && \la Q_{ij} \ra = 4 C_E \la E_{ij} \ra \,.
\el
The substitution rules for converting the action to amplitudes is~\cite{Chung:2018kqs}
\bl
\bld
\partial_\m h_{\a\b} \to - i k_\mu \ve_{\a\b} (k) \,,
&& u^\m \to p_1^\m / m \,,
\eld
\el
where $\ve_{\a\b} (k)$ is the polarization tensor of the graviton with momentum $k^\m$. The overall scaling factor $N = 2m$ is again fixed by comparing to the on-shell three-point amplitude.
\bl
\bld
\CA_{mmh}^{\text{Eff. Act.}} &= - \frac{\k N}{2m} \ve_{\m\n} p_1^\m p_1^\n \,,
\\ \CA_{\bar{\f} {\f} h}^{\text{Scalar}} &=  \k \ve_{\m\n} p_1^\m p_2^\n \,.
\eld
\el
This results in the on-shell amplitude modifications
\bl
\bld
\D \CA_{mhhm}^{++} &= \frac{\k^2 (C_E - C_B) m [23]^4}{16} \,,
\\ \D \CA_{mhhm}^{+-} &= \frac{\k^2 (C_E + C_B) \sbra{2}p_1\ket{3}^4}{16 m^3} \,.
\eld \label{eq:GravComptonEff} 
\el
Matching \eqc{eq:GravComptonContact1} to \eqc{eq:GravComptonEff} fixes the Love numbers as
\bl
C_E =  \left( \frac{4\tilde\CC}{m} \right)^2 \left[\frac{1}{\D m} + \frac{2}{m} \right] \,,\, C_B = - \left( \frac{4\tilde\CC}{m} \right)^2 \frac{2}{m} \,. \label{eq:GravsusceptQFT} 
\el
Again, magnetic Love number $C_B$ is suppressed compared to its electric counterpart $C_E$ by the ratio ${C_B}/{C_E} \simeq - {2\D m}/{m}$. Combining \eqc{eq:GravDR} with \eqc{eq:GravsusceptQFT} yields
\bg 
\frac{\G_{m' \to mh}}{C_E} =  \frac{\k^2 \omega^5 \D m}{4 \pi} =  8 G \omega^6 \,, \label{eq:gravdecay2pol} 
\eg
in the nonrelativistic approximation. The result \eqc{eq:gravdecay2pol3} is obtained by summing over other decay channels as in the electromagnetic case \eqc{eq:decay2pol3},
\bl
C_E &=  \frac{1}{8G} \left[ \sum_{m'>m} \frac{\G_{m' \to mh}}{(\D m)^6} - \sum_{m'<m} \frac{\G_{m \to m'h}}{(\D m)^6} \right] \,. \nn 
\el
The factor of $G^{-1}$ in the above formula is expected to be canceled by $\k^2 \propto G$ appearing in the decay rates $\G$, resulting in the overall scaling $G^0$. This is consistent with the expectation that tidal responses are properties of the material itself, so it should not depend on Newton's constant $G$. 

\section{Quantum Love number of SBHs}
An interesting prediction of classical general relativity in four spacetime dimensions is that SBHs have exactly vanishing Love numbers~\cite{Damour:2009vw,Binnington:2009bb,Kol:2011vg,Chakrabarti:2013lua,Gurlebeck:2015xpa,Porto:2016zng,Hui:2020xxx,LeTiec:2020bos,Charalambous:2021mea,Charalambous:2021kcz}. 
The sum rule \eqc{eq:gravdecay2pol3} can be used to check if this property persists at the quantum level. 

The relevant transition rates for the sum rule \eqc{eq:gravdecay2pol3} are emission channels of gravitons as $s$-waves ($l=0$ or $l=2$ depending on conventions), which may consist of horizon excitations (quasinormal modes) and Hawking radiation.
The contributions from the latter will be computed in this section. While the contributions from the former are also expected to be computable, as quasinormal modes have already been studied extensively in the literature~\cite{Kokkotas:1999bd,Berti:2009kk}, it is unclear whether this would be double-counting emission channels as black holes with horizon excitations of a single quantum would be indistinguishable from an unexcited black hole at the classical level.

Restoring $\hbar$ and $c$ to \eqc{eq:gravdecay2pol3} leads to the expression~\footnote{J.-W.K. would like to thank Lance Dixon, Callum Jones, and Julio Parra-Martinez for motivating this dimensional analysis.}
\bl
C_E &=  \frac{c^5}{8G} \left[ \sum_{m'>m} \frac{\G_{m' \to mh}}{\omega^6} - \sum_{m'<m} \frac{\G_{m \to m'h}}{\omega^6} \right] \,, \label{eq:sumrule_dim} 
\el
where $\omega = \abs{\D m} c^2 / \hbar$ is the frequency of the emitted graviton and dimensions of $C_E$ and $\G$ were set as $[ML^2T^{2}]$ and $[T^{-1}]$. At first glance, the contributions from Hawking radiation do not seem to be quantum suppressed as Hawking emission rate $d\G / d\omega$ \eqc{eq:HRemdens} does not scale with $\hbar$. This seems to be at odds with the conclusion of Ref.\cite{Goldberger:2019sya} that Hawking radiation does not contribute to classical observables~\footnote{The authors are grateful to Rafael Porto and an anonymous referee for bringing this reference to our attention.}, unless the ``classical" contributions from Hawking radiation are delicately canceled. This indeed is the case; because of high suppression due to $\omega^{-6}$ it may be assumed that the dominant contribution comes from small $\omega$, and Hawking radiation's contribution to the sum \eqc{eq:sumrule_dim} can be approximated as
\bl
C_{E}^{\text{HR}} &=  \frac{c^5}{8G} \sum_{\omega>0} \frac{\Gamma(M+\hbar c^{-2} \omega ; \omega) - \Gamma(M; \omega)}{\omega^6} 
\,, \nn 
\el
which can be computed using the integral
\bl
C_{E}^{\text{HR}} &=  \frac{c^3 \hbar}{8G} \int_0^\infty \frac{d\omega}{\omega^5} \frac{\partial}{\partial M} \left[ \frac{d \G}{d\omega} (M;\omega) \right] \,, \label{eq:HRloveint} 
\el
where $d\G / d\omega (M; \omega)$ is the total Hawking emission rate density of a SBH of mass $M$ to emit a single graviton of frequency $\omega$ in the $s$-wave channel. 
For $G M\omega \ll 1$, the emission rate is \cite{Page:1976df} 
\bl
\bld
\frac{d \G}{d\omega}
&= \frac{16}{45 \pi^2} \frac{A M^4 \omega^6}{e^{2 \pi \omega / \k} - 1}
= \frac{256}{45 \pi} \frac{(G/c^3)^6 M^6 \omega^6}{e^{8 \pi G M \omega / c^3} - 1} \,.
\eld \label{eq:HRemdens}
\el
Assuming gravitons of Hawking radiation are only emitted through single quantum emission processes, the emission rate density \eqc{eq:HRemdens} fully contributes to the sum rule \eqc{eq:gravdecay2pol3}. Inserting \eqc{eq:HRemdens} into \eqc{eq:HRloveint} leads to a finite value for the Hawking radiation induced Love number~\footnote{The contributions from $G M\omega \gg 1$ channels are overestimated since the greybody factor is much smaller. However, this error is expected to be neglegible due to high suppression from $\omega^{-6}$ as explained in the main text.}
\bl
C_{E}^{\text{HR}} &=  \frac{G^3 M^3 \hbar}{135 \pi c^9} \,, \nn 
\el
which was given as Eq. \eqc{eq:HRlove} in the beginning of this paper. The spin corrections $J = 2\hbar$ for the channels $m'>m$ are irrelevant at this order in $\hbar$. This result is Planckian suppressed compared to the ``natural" scale of tidal Love number for astrophysical black holes~\footnote{Normalization for $k_2$ follows from $Q_{ij} = - \frac{2}{3} k_2 \frac{R^5}{G} \partial_i \partial_j \Phi$.}
\bl
k_2 &= \frac{3 \times C_{E}^{\text{HR}}}{M R^2 (R/c)^2} 
=  \frac{1}{90} \left( \frac{m_{\text{Pl}}}{M} \right)^2 \,, \label{eq:HRlovePlanck} 
\el
where $R = 2GM/c^2$ is the Schwarzschild radius and $m_{\text{Pl}} = \sqrt{\hbar c / 8\pi G}$ is the reduced Planck mass. Extension to quantum corrections of electric polarizability for SBHs, of which its classical value has been computed in refs.\cite{Damour:2009va,Hui:2020xxx,Charalambous:2021mea}, would be straightforward.

\section{Discussions}
The result \eqc{eq:gravdecay2pol3} is a robust prediction of perturbative quantum gravity on a flat background in four spacetime dimensions, in the sense that only kinematical conditions were used to arrive at the result~\footnote{In fact, the sum rule should be derivable within nonrelativistic QM by slightly modifying the derivation of quantum mechanical Love number in Ref.\cite{Brustein:2020tpg}. The sign difference of the Love number comes from the sign difference of the interaction potential $V_{int} = \pm \half Q_{ij} E_{ij}$. The validity of the derivation is dependent on two standard assumptions of nonrelativistic QM, however; the applicability of time-independent perturbation theory and Fermi's golden rule for transition rates.
}. Apart from the implicit assumption of nondegenerate mass spectrum, this result only relies on the fact that gravitons are massless spin-2 particles, which is a shared feature among a large class of quantum gravity models. Generalization of the result \eqc{eq:gravdecay2pol3} to higher induced multipole moments can be done in an analogous fashion, although the spin of intermediate particles and exponents of the mass difference would have to be changed.

When deriving the quantum correction \eqc{eq:HRlove}, it was assumed that the semiclassical Hawking radiation spectrum reliably captures the numerators in the sum rule \eqc{eq:gravdecay2pol3}. Note that this assumption includes the assumption that all gravitons emitted as Hawking quanta are emitted through single quantum emission channels~\footnote{The thermal emission spectrum \eqc{eq:HRemdens}, which is nonunitary, might not be suitable for application to \eqc{eq:gravdecay2pol3} which was derived assuming unitarity. The authors would like to thank Sang-Heon Yi for discussions on this point.}, which is coherent with our usual understanding of Hawking radiation; an entangled pair is created near black hole horizon and one of the pair falls into the black hole, while the other propagates out to infinity and gets picked up by a piece of measurement apparatus as radiation~\cite{Mathur:2011uj,Almheiri:2020cfm}. 
Therefore a large violation of the prediction \eqc{eq:HRlove} could be tied to the assumption that Hawking quanta are predominantly created through single quantum emission processes, signaling a flaw in our picture of Hawking radiation as a gravitational version of Schwinger pair creation.
This may have connections to the information problem and entanglement properties of black holes~\cite{Hawking:1976ra,Page:1993df,Page:1993wv,Almheiri:2012rt,Bae:2020lql}.

Although the high suppression of quantum corrections of the tidal Love number compared to its typical scale \eqc{eq:HRlovePlanck} renders experimental observation of the correction for astrophysical black holes unlikely in the near future, the prediction \eqc{eq:HRlove} can be tested when given a model of quantum black holes. In this sense, the sum rule \eqc{eq:gravdecay2pol3} and its application to Hawking radiation \eqc{eq:HRloveint} could be used as a consistency check for models of realistic quantum black holes describing Hawking radiation~\footnote{The sum rule \eqc{eq:gravdecay2pol3} would apply if the black hole is considered to be in a pure state with a definite mass eigenvalue. Generalization to cases involving superposition states of differing mass eigenvalues will be left for future work.}. A peculiar feature of this prediction is that quantum corrections to the tidal Love number seems to be sensitive to \emph{infrared} physics of black holes in contrast to the expectation that it is sensitive to ultraviolet physics~\cite{Porto:2016zng}, as the sum rule \eqc{eq:gravdecay2pol3} is sensitive to the soft side of the Hawking spectrum. In this light, the quantum corrections to the tidal Love number \eqc{eq:HRlove} could be interesting for recent discussions on soft hairs of black holes~\cite{Hawking:2016msc,Mirbabayi:2016axw,Haco:2018ske,Nomura:2019qps,Haco:2019ggi,Averin:2019zsi,Donnay:2020fof,Pasterski:2020xvn,Flanagan:2021ojq}.

One conceptual loophole is that decaying massive particles were involved in computing the scattering amplitudes, or the $S$-matrix elements. This means unstable particles were included in the spectrum of asymptotic states. However, since the electromagnetic version \eqc{eq:decay2pol3} matched exactly to the results \eqc{eq:decayQM} from a well-known solvable model and nonrelativistic quantum mechanical derivation of the sum rule is likely to exist~\footnote{See the footnote on first paragraph of discussions.}, this conceptual subtlety is likely to be inconsequential. Also, it would be interesting to investigate the consequences of including the decay width to the $m'$ propagator appearing in \eqc{eq:Compton} and \eqc{eq:ComptonGrav}; $(q^2 - m'^2)^{-1} \to (q^2 - m'^2 + i m' \G)^{-1}$.


\section{Acknowledgements}
The authors would like to thank Yu-Tin Huang, Sang-Heon Yi, Dong-han Yeom, and Sangmin Lee for insightful discussions. J.-W.K. would also like to thank Andreas Brandhuber, Gabriele Travaglini, Costis Papageorgakis, Chris White, Manuel Accettulli Huber, and Stefano De Angelis for helpful discussions, and \emph{QCD meets gravity VI} workshop for providing an arena of stimulating discussions. M.S. would like to thank Nakwoo Kim for instructive discussions. J.-W.K. was supported by the Science and Technology Facilities Council (STFC) Consolidated Grant No. ST/T000686/1 \emph{``Amplitudes, Strings and Duality”}. M.S. was supported in part by National Research Foundation (NRF) of Korea grant  No. NRF-2019R1A2C2004880 and No. NRF-2018R1A2A3074631, and in part by a scholarship from Hyundai Motor Chung Mong-Koo Foundation. 

\bibliography{ref.bib}

%

\end{document}